\documentclass[english]{emulateapj}

\usepackage{textcomp}
\usepackage{amsmath,amssymb,amsfonts}
\usepackage{hyperref}
\usepackage{txfonts}
\usepackage{subfigure}
\usepackage{natbib}

\begin{document}

 \title{\bf{Post-main-sequence evolution of icy minor planets. III. water retention in dwarf planets and exo-moons and implications for white dwarf pollution}}

\author{Uri Malamud and Hagai B. Perets}
\affil{Department of Physics, Technion, Israel}
\email{uri.mal@tx.technion.ac.il ~~~~ hperets@physics.technion.ac.il}

\newpage

\begin{abstract}
Studies suggest that the pollution of white dwarf (WD) atmospheres arises from the accretion of minor planets, but the exact properties of polluting material, and in particular the evidence for water in some cases are not yet understood. Previous works studied the water retention in minor planets around main-sequence and evolving host stars, in order to evaluate the possibility that water survives inside minor planets around WDs. However, all of these studies focused on small, comet-sized to moonlet-sized minor planets, when the inferred mass inside the convection zones of He-dominated WDs could actually also be compatible with much more massive minor planets. In this study we therefore explore for the first time, the water retention inside exo-planetary dwarf planets, or moderate-sized moons, with radii of the order of hundreds of kilometres. We now cover nearly the entire potential mass range of minor planets. The rest of the parameter space considered in this study is identical to that of our previous study, and also includes multiple WD progenitor star masses. We find that water retention in more massive minor planets is still affected by the mass of the WD progenitor, however not as much as when small minor planets were considered. We also find that water retention is now almost always greater than zero. On average, the detected water fraction in He-dominated WD atmospheres should be at least 5\%, irrespective of the assumed initial water composition, if it came from a single accretion event of an icy dwarf planet or moon. This finding also strengthens the possibility of WD habitability. To finalize our previous and current findings, we provide a code which may be freely used as a service to the community. The code calculates ice and water retention by interpolation, spanning the full mass range of both minor planets and their host stars. 
\end{abstract}


\section{Introduction}\label{S:Intro}
Despite the typically short sinking time scale of elements heavier than helium in the atmospheres of WDs \citep{Koester-2009}, between 25\% to 50\% of all WDs \citep{ZuckermanEtAl-2003,ZuckermanEtAl-2010,KoesterEtAl-2014} are found to be polluted with heavy elements. This is suggestive of their ongoing accretion of planetary material \citep{DebesSigurdsson-2002,Jura-2003,KilicEtAl-2006,Jura-2008}, originating from minor planets that survive the main sequence, red giant branch (RGB) and asymptotic giant branch (AGB) stellar evolution phases, and remain bound to the WD. When some of these minor planets are perturbed to orbits with proximity to the WD, they are thought to be tidally disrupted and form a circumstellar disk \citep{VerasEtAl-2014a,VerasEtAl-2015}, which eventually accretes onto the WD \citep{Jura-2008,Rafikov-2011,MetzgerEtAl-2012}.

The spectroscopic analysis of WD atmospheres \citep{WolffEtAl-2002,DufourEtAl-2007,DesharnaisEtAl-2008,KleinEtAl-2010,GansickeEtAl-2012} as well as infra-red spectroscopy of the debris disks themselves \citep{ReachEtAl-2005,JuraEtAl-2007,ReachEtAl-2009,JuraEtAl-2009,BergforsEtAl-2014}, are typically consistent with 'dry' compositions, characteristic of inner solar system objects. Studies that analysed the atmospheric composition of large samples of nearby He-dominated WDs, inferred their collective water mass fraction to be of order $\sim$1\%, and no more than a few \% even for particular subsets \citep{JuraXu-2012,JuraYoung-2014,PietroGentileFusilloEtAl-2017}. On the other hand, an increasing number of individual WDs \citep{FarihiEtAl-2013,RaddiEtAl-2015,XuEtAl-2017} have now been inferred to contain a large water mass fraction, ranging between 26\%-38\%. Polluted WDs are therefore the only current means to understand exo-planetary composition. However, linking the inferred quantity of water in WD atmospheres to the origin of polluting planetary material is not trivial, and requires additional assumptions and calculations. 

Several studies have previously focused on water retention inside minor planets, as they evolve through and off the main-sequence of their host stars. Understanding how water survive inside minor planets as a function of their intrinsic characteristics, is a necessary first step in beginning to understand the origin and nature of polluting planetary material. The seminal work of \cite{SternEtAl-1990} introduced a simple sublimation model that considered the initial orbital distance and size of minor planets as they evolve around a post-main-sequence 1 M$_{\odot}$ star. \cite{JuraXu-2010} considered a more advanced model, taking into account conductive heat transport to the interior, and also the minor planets' orbital expansion during the post-main-sequence stellar evolution of a 1 M$_{\odot}$ and 3 M$_{\odot}$ host stars. \cite{MalamudPerets-2016} (hereafter MP16) utilized a modern and more sophisticated code that consistently accounts for the thermal, as well as physical (rock/water differentiation), chemical (rock hydration and dehydration) and orbital evolution of icy minor planets. Contrary to previous models, it starts at the beginning of the main-sequence (of a 1 M$_{\odot}$ host star), and considers new parameters: the formation time of the minor planet (which affects its early evolution through radiogenic heating and thus its internal differentiated state) and also variation in the initial minor planet composition. \cite{MalamudPerets-2017} (hereafter MP17) finally investigated multiple host star masses, ranging from 1 M$_{\odot}$ to 6.4 M$_{\odot}$ (corresponding to $\sim$0.5-1 M$_{\odot}$ WDs), as well as various host star metallicities.

Until now all of these studies shared a common feature - they were all limited to relatively small minor planets, ranging from comet-sized to minor planets with radii of about 100 km or slightly more. This has been justified by the fact that small minor planets are far more numerous and are likely to trigger most of the pollution in WDs. Nevertheless, intermediate and large minor planets are far more massive. The occasional accretion of even a single minor planet of this size may be sufficient in order to account for all the mass in the convection zone of He-dominated WD atmospheres, whereas for small minor planets multiple accretion events are typically needed in order to accumulate and reach the required mass (e.g., see Figure 6 in \citep{Veras-2016}). Indeed, it should be kept in mind that we currently do not know exactly what path or paths typically lead to accretion. Perturbed minor planets could be of various sizes, locations, potential system configurations and perturbation mechanisms \citep{DebesSigurdsson-2002,BonsorEtAl-2011,DebesEtAl-2012,KratterPerets-2012,PeretsKratter-2012,ShapeeThompson-2013,MichaelyPerets-2014,VerasGansicke-2015,StoneEtAl-2015,HamersPortegiesZwart-2016,Veras-2016,PayneEtAl-2016,PetrovichMunoz-2016,CaiazzoHeyl-2017,PayneEtAl-2017,StephanEtAl-2017}. Most of the aforementioned mechanisms do not place specific limits on their size, and some mechanisms even discuss particular objects such as liberated exo-moons \citep{PayneEtAl-2016,PayneEtAl-2017} which have a higher chance of being massive. While the actual hunt for exo-moons is still at its infancy stage \citep{TeacheyEtAl-2017}, knowledge from our own solar system shows this to be the case. 

The goal of this paper is therefore to complement all previous studies by calculating, for the first time, the fate of water inside moon-sized or dwarf-planet-sized objects. Unlike small minor planets, which tend to have a low bulk density and thus very large porosity (we refer to the discussion in Section 2.1 (d) and Table 1 in MP16), the objects discussed in this paper will require a different treatment of porosity, since minor planets of this size evolve mechanically due to the combination of their large internal pressures by self-gravity and also high temperatures, which tend to reduce or eliminate porosity completely \citep{MalamudPrialnik-2015}. In what follows we briefly outline in Section \ref{S:Model} the size range investigated and the model used in this study. In Section \ref{S:Results} we initially present one evolutionary path for a dwarf-planet-sized minor planet and emphasize the differences to previous studies. We then present the water retention results for the entire parameter space. We discuss the results in Section \ref{S:discussion} and finalize our work in this and previous papers by providing the community with a code that calculates water retention, spanning the entire minor planet mass spectrum.

\section{Model}\label{S:Model}
In the framework of this paper, the terms "intermediate", "medium" or "moderate" are used with reference to objects which are large enough to be in hydrostatic equilibrium, so any object with a radius smaller than about 200 km can be excluded from this definition \citep{LineweaverNorman-2010}, but are not so large as to permit the occurrence of high-pressure phases of water ice since our model currently only treats the low-pressure 'Ice I' phase. This limits the object radius to be of the order of several hundreds of km, although the precise size determination depends on various parameters, like the assumed initial composition, structure, temperature etc. For example, if one assumes an initially differentiated structure, the self-gravity pressure in the icy mantle will be compatible with the Ice I phase, even for a dwarf planet the size of Pluto (radius$\sim$1200 km). However if one initially assumes a homogeneous ice-rock structure prior to differentiation, as we do in our model (in order to calculate differentiation as a function of variable formation times), the self-gravity pressure in Pluto's core is initially above that of Ice I. Given this constraint of initial homogeneity, objects that are much larger than about 600-700 km in radius cannot be considered in the framework of this paper, and their analysis remains the goal of future studies. 

As we shall see, their exclusion from this paper is of very little significance, at least from an observational point of view, as follows. At our current level of understanding the distribution of mass, when observed in WD atmospheres (see Figure 6 in \citep{Veras-2016}) follows a bell shape with a peak at around $\sim$$10^{22}$ g. Previous studies therefore investigated about half this mass range, ranging from well below the minimum and reaching approximately the peak in the mass distribution. This study will roughly cover the other half, since beyond a mass of $\sim$$10^{24}$ g, just under the limit of this study, the tail becomes relatively unimportant. 

Our evolution model couples the thermal, physical, chemical, mechanical and orbital evolution of icy minor planets of various sizes. It considers the energy contribution primarily from radiogenic heating, latent heat released/absorped by geochemical reactions and surface insolation. It treats heat transport by conduction and advection, and follows the transitions among three phases of water (crystalline ice, liquid and vapor), and two phases of silicates (hydrous rock and anhydrous rock). In our predecessor studies on water retention and WD pollution (MP16,MP17) we considered only small minor planets, and therefore we deactivated a specific feature in the model that deals with internal mechanical changes. In this work we add an additional level of complication, and maintain hydrostatic equilibrium. This involves continuously solving the hydrostatic equation, using an equation of state that provides the density to pressure relation of a porous mix of rock/ice. For details see equtions 1 and 16 in \cite{MalamudPrialnik-2015} and references therein. 

All other model details, including equations, parameters and numerical scheme, are identical to our previous work and can be reviewed in MP16. Our code requires as input the stellar evolution of WD progenitor stars of different masses (that is, change in star luminosity and mass as a function of time). These inputs are also compatible with our previous work. See MP17 for details on how they were obtained using the MESA stellar evolution code. We consider progenitor masses of 1, 2, 3, 3.6, 5 and 6.4 M$_{\odot}$, with a metallicity of 0.0143 (or [Fe/H]=0 -- the typically used iron abundance relative to solar) corresponding to the final WD masses of 0.54, 0.59, 0.65, 0.76, 0.89 and 1 M$_{\odot}$ respectively. 

The final outcome of the main-sequence, RGB and AGB stellar evolution phases in terms of minor planet water retention, depends on five different parameters. The first parameter is the progenitor's mass. More massive progenitors correlate with a more luminous stellar evolution, albeit a shorter lifetime and also a higher initial (progenitor) to final (WD) mass ratio. The former affects water retention negatively, while the latter two have a positive effect. We also investigate four characteristic variables related to the minor planets: their size (mass/radius), initial orbital distance, formation time and initial rock/ice mass ratio. To comply with our previous work, we consider a similar parameter space, however in this study larger minor planet masses and thus radii are considered, as well as additional changes in the orbital distances and formation times, as outlined below. 

(1) Object mass/radius - as previously mentioned, past studies investigated the minor planet mass range up to $10^{22}$ g. This paper aims to investigate the $10^{23}$-$10^{24}$ g mass range. For the purpose of familiarity, we consider the mass of two well-known Solar system objects, Enceladus and Charon, as our reference masses. Enceladus has a mass of $\sim$$10^{23}$ g whereas Charon has a mass of $\sim1.5$$\cdot10^{24}$ g. These two objects therefore differ in mass by roughly one order of magnitude, and span the desired mass range almost perfectly. In terms of size, the radius is \emph{not} constant as in previous studies. Our previous studies did not take into account mechanical changes, and therefore porosity never decreased. Whenever water migrated toward the surface or sublimated off the surface, the initial porosity simply increased, and thus the radius was fixed, while mass wasn't necessarily. Here the case is different, since massive minor planets undergo huge mechanical, structural and compositional changes that considerably alter their size. For example, given a fixed mass, the initial radius may depend on the value chosen for the minor planet's bulk density, and also on the assumed initial rock/ice mass ratio. More rocky objects will have a smaller initial radius whereas more icy objects a larger initial radius. The radius may increase over time as water initially melts, migrates and then freezes to form a differentiated internal structure. The radius may also decrease over time as change in temperature or pressure by self-gravity will diminish internal porosity, or as the body looses water due to insolation from the star. It is therefore the case here, that neither the initial mass nor the initial radius are fixed. Nevertheless, it is convenient to discuss a \emph{characteristic} radius, which is approximately the radius of the minor planet at the beginning of its evolution, before insolation from the star starts to expel mass. For the two reference masses chosen, the characteristic radii will be approximately 270 and 600 km respectively. Naturally, if and when a minor planet expels water via insolation, the radius can decrease well below this characteristic value (up to 26\% less). We note that size changes in minor planets can naturally occur even around non-evolving stars (e.g., see \cite{MalamudEtAl-2017}), but these changes are amplified when insolation or other processes (as in \cite{MalamudPrialnik-2016}) lead to massive ablation of ice.

(2) Orbital distance - we consider a range of possible initial orbital distances (note that with stellar mass loss the orbit undergoes expansion as the minor planet conserves its angular momentum). The minimal initial orbital distance is 3 AU. Below approximately 3 AU, a massive planet (and by extension also its moons) runs the risk of being engulfed or otherwise tidally affected by the expanding envelope of the post-main sequence RGB \citep{KunitomoEtAl-2011,VillaverEtAl-2014} or AGB \citep{MustillVillaver-2012} star. Another consideration for icy minor planets is the location of the snowline \citep{KennedyKenyon-2008}, although minor planets could potentially also migrate inward from their initial birthplace. Overall, a minimum of $\sim$3 AU was chosen as the minimal distance. The maximum orbital distance changes from 75 AU to 200 AU, depending on the progenitor star's mass. It is determined according to the water retention upper bound, defined as the distance at which full water retention is ensured. The water retention upper bound was previously found to increase as a function of the progenitor's mass (MP17). In this study it behaves in the same way, and therefore the number of grid points increases slightly in order to cover a wider distance range.

(3) Formation time - the formation time of a minor planet is defined as the time it takes a minor planet to fully form, after the birth of its host star. Since here we only consider first-generation minor planets, this time is usually on the order of $\sim 10^0-10^1$ Myr. The formation time determines the initial abundance of short-lived radionuclides, and thus the peak temperatures (hence, internal structure) attained during its early thermal evolution. Although it is clear that the formation time also depends on the orbital distance, the exact relation is unconstrained, which is why we set the formation time as a free parameter. We consider the following formation times: 3, 4, 5 and 10 Myr. This choice is compatible with our previous works, however here we include an additional formation time of 10 Myr, since in more massive minor planets the internal temperature can build up more easily. At approximately 10 Myr formation time, short-lived radionuclides decay so much that effectively only long-term radiogenic heating becomes important. The initial abundances of radionuclides are assumed to be identical to the canonical values in the solar system, for lack of a better assumption.

(4) Initial rock/ice mass ratio - this ratio initially depends on the location of the object as it forms in the protostar nebula, and like the formation time this parameter is unconstrained (see MP16 for various estimations). We consider three initial rock/ice mass ratios to allow for various possibilities: 1, 2 and 3 (that is, a rock mass fraction of 50\%, 67\% and 75\% respectively), complying with previous work.

The number of models for a single stellar evolution is thus determined by the number of variable minor planet parameters (2 x (7-9) x 4 x 3). Since we have six different progenitor star masses, we have well over one thousand production runs in total. These models were calculated using a cluster computer. The typical run time of each model was on the order of several hours on a single 2.60GHz, Intel CPU. All other model parameters are equal, and identical to the parameters used in Table 2 of MP16.

\section{Results}\label{S:Results}
\subsection{The evolutionary course}\label{SS:Evolution}
\begin{figure}
	\begin{center}
		\subfigure[Temperature]
		{\label{fig:Temperature}\includegraphics[scale=0.5]{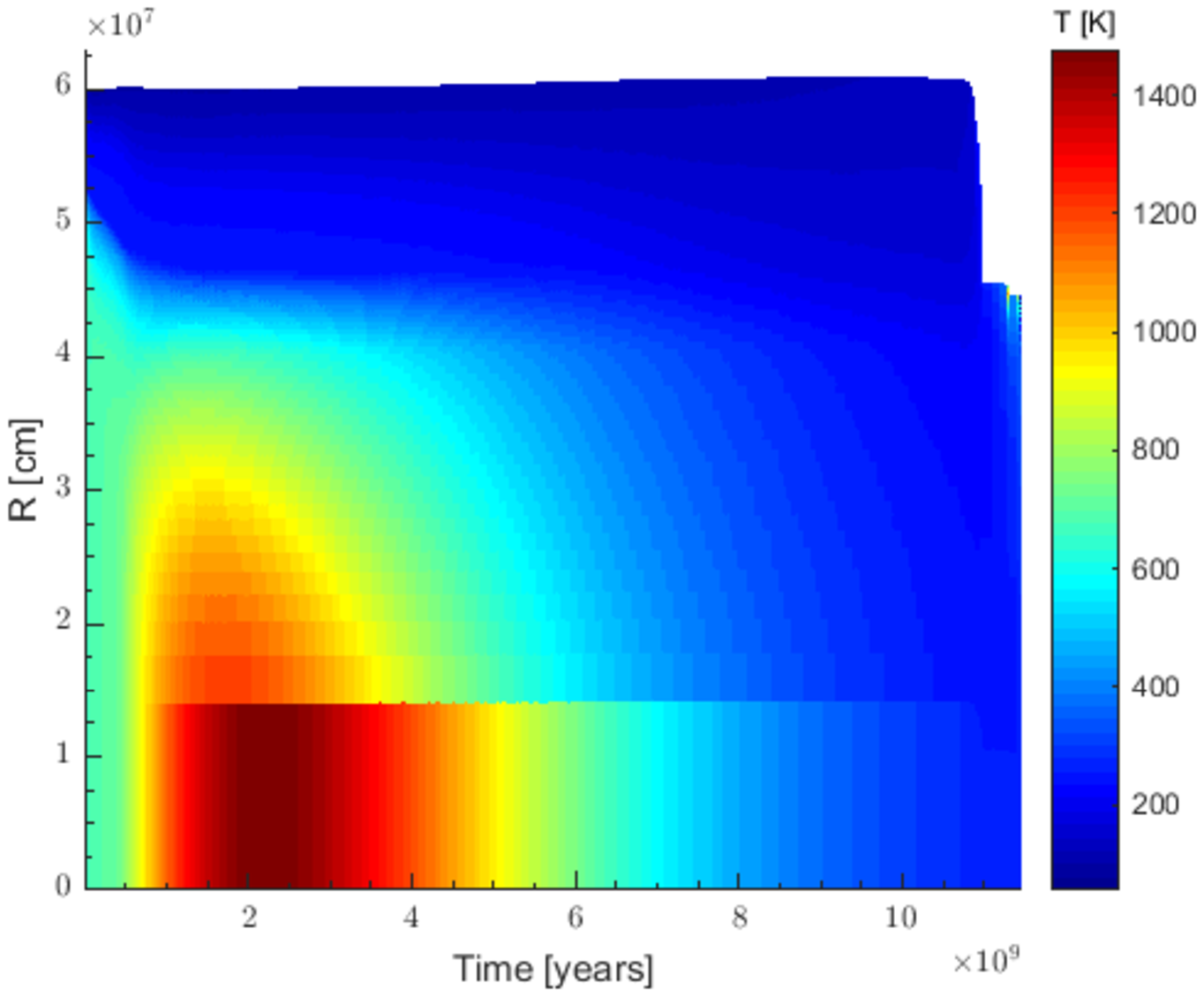}}
		\subfigure[Water fraction]
		{\label{fig:WaterFraction}\includegraphics[scale=0.5]{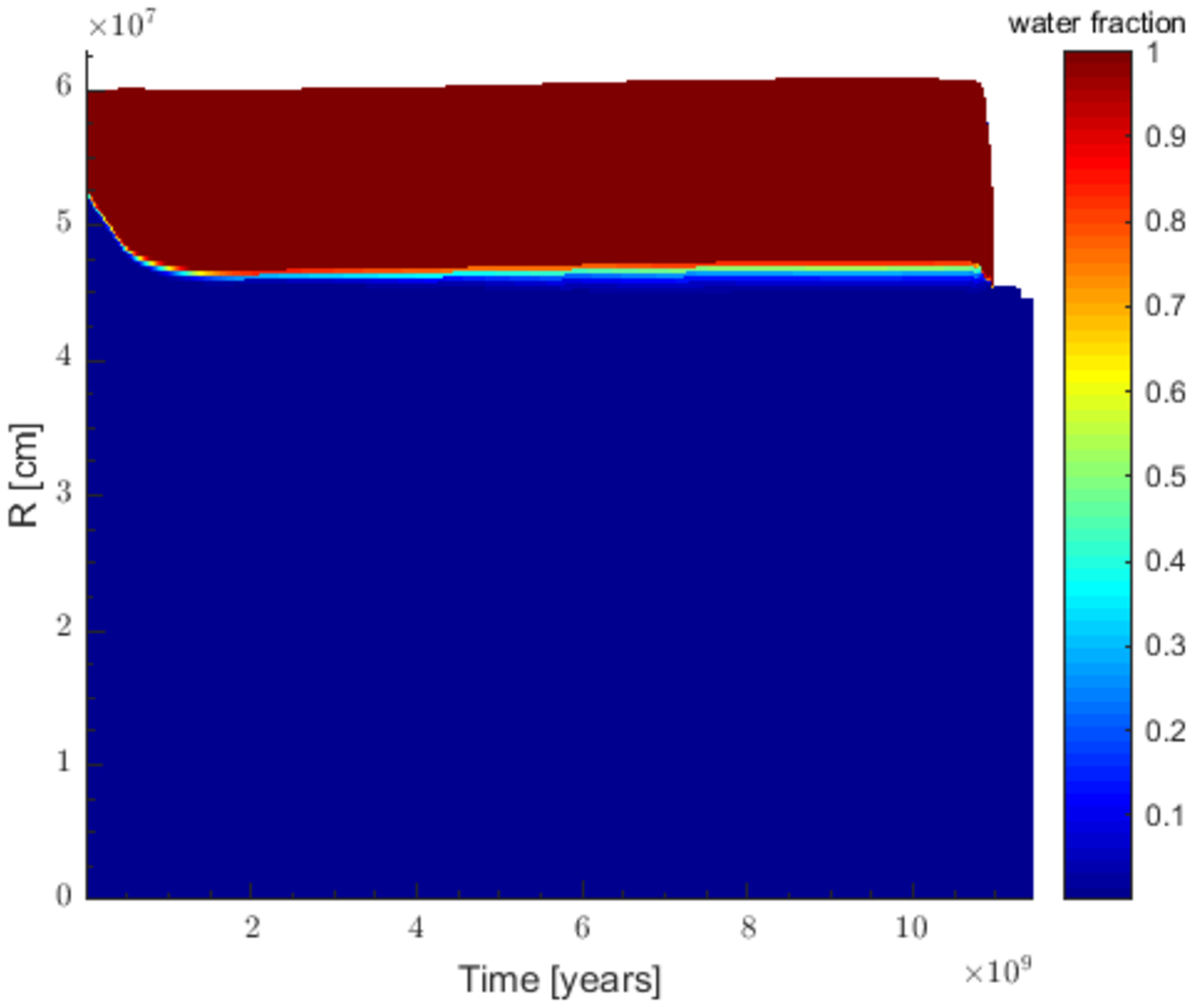}}
	\end{center}
	\caption{A surface plot depicting the evolution of (a) temperature and (b) water fraction, as a function of time (x-axis) and radial distance from the center of a minor planet (y-axis). The minor planet shown has an initial radius of $\sim$600 km, a rock fraction of 75\%, a formation time of 3 Myr and it orbits a 1 M$_{\odot}$ progenitor star at an initial orbital distance of 3 AU. The evolution duration is 11.46 Gyr, from the beginning of the main sequence to the formation of the WD.}
	\label{fig:Evolution}
\end{figure}

As mentioned in Section \ref{S:Intro}, here we present a single detailed example for an evolutionary calculation. Our goal is to highlight the differences between the evolution of small minor planets and that of larger, dwarf-planet-sized minor planets. In Figure \ref{fig:Evolution} we show the entire evolution for a minor planet with the largest size considered in our sample (radius=$\sim$600 km) at an initial orbital distance of merely 3 AU. This dwarf planet is 75\% rock by mass, its assumed formation time is 3 My and its host star has the mass of 1 M$_{\odot}$ and the metallicity of [Fe/H]=0. We choose this particular combination of parameters as an example since it represents a minor planet that undergoes the most extreme conditions possible in our sample and attains very high surface and internal temperatures for the longest period of time. It is also a familiar and intuitive example, since we are considering an object only slightly more massive than Ceres, and at a similar orbital distance to Ceres. Its evolution around an almost sun-like star may therefore be seen as a exo-planetary Ceres-analog, or to be exact, a Charon-analog placed at 3 AU (since Ceres has a more rocky composition and higher bulk density than Charon). The final orbital distance (that is, around the WD) after stellar mass loss is 5.55 AU.

The course of the evolution is illustrated by two surface plots showing the temperature \ref{fig:Temperature} and the relative fraction of water \ref{fig:WaterFraction} as a function of time and radial distance from the centre of the body. In both Panels, the x-axis shows the time interval, ranging from 0 to 11.46 Gyr. The y-axis shows the radial distance from the centre of the body. Note that the upper boundary of the y-axis changes with time. It increases as the inner rocky core dehydrates due to rising internal temperatures (see Panel \ref{fig:Temperature}), and the icy mantle thickens by about $\sim$7-8 km due to water released from the underlying rock (Panel \ref{fig:WaterFraction}). After this gradual size increase, at around 10 Gyr the size slowly starts to decrease as a rise in surface temperature triggers sublimation. By 11 Gyr the icy mantle sublimates entirely. The radius then reaches $\sim$450 km, and it only reduces a little bit further during the AGB phase when extremely high surface temperatures compress the near-surface rock porosity, and also alter the rock's composition via dehydration.

\begin{figure}
	\begin{center}
		\label{fig:CS}\includegraphics[scale=1]{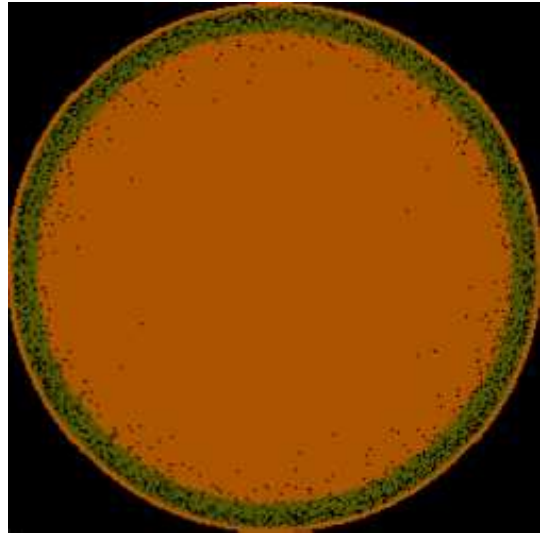}
		\caption{Animated figure (duration - 30 s) featuring the evolution of a 600 km radius minor planet around a 1 M$_{\odot}$ progenitor star at 3 AU, with a 3 Myr formation time and 75\% initial rock fraction. Colour interpretation: {\it black} (pores); {\it white} (water ice); {\it blue} (liquid water); {\it brown} (anhydrous rock); {\it olive} (hydrated rock); and {\it red} (molten rock). The animation shows the following sequence of processes: (a) differentiation of an initially homogeneous body into a hydrous rocky core underlying a thin ice-enriched crust; (b) dehydration and compression of the rocky core while icy mantle thickens; (c) gradual sublimation of icy mantle via rising surface temperature; and (d) further rise in surface temperature leading to near-surface rock dehydration and compaction.}
	\end{center}
\end{figure}

Fig. \ref{fig:CS} shows the compositional cross-section of this minor planet as it evolves (since this is an animated figure, one may view the evolution, in addition to the end state which is depicted by the still image). It easily illustrates the large differences in water retention between this study and previous studies, as follows: 1) In small minor planets the dehydration of the rocky core was marginal at best (MP16), amounting to no more than a 5\% of the rock and typically none at all. Here however we see that dehydration can have a huge effect. Almost the entire rock mass becomes dehydrated by the evolution's completion, and only a small fraction remains hydrated in a thin shell close to the surface. 2) In small minor planets the porosity remains high and even increases if water ice sublimates out of the interior (see Figs. 2-4 in MP17). Here however porosity decreases considerably, and in some cases it can even reduce to zero in the rocky core. Porosity is very important because it greatly decreases the effective thermal conductivity \citep{Smoluchowski-1981}, and thus, heat transport. In small minor planets the near-surface rock is not always hydrated to begin with, and even if it is, dehydration of external rock during the AGB phase is almost negligible since heat penetrates the interior much more slowly, as a result of the high porosity.

\subsection{Water retention}\label{SS:Retention}

\begin{figure*}[h!]
	\begin{center}
		\subfigure[1 M$_{\odot}$ progenitor - 0.54 M$_{\odot}$ WD - 11.4 Gyr evolution] {\label{fig:1sm}\includegraphics[scale=0.5]{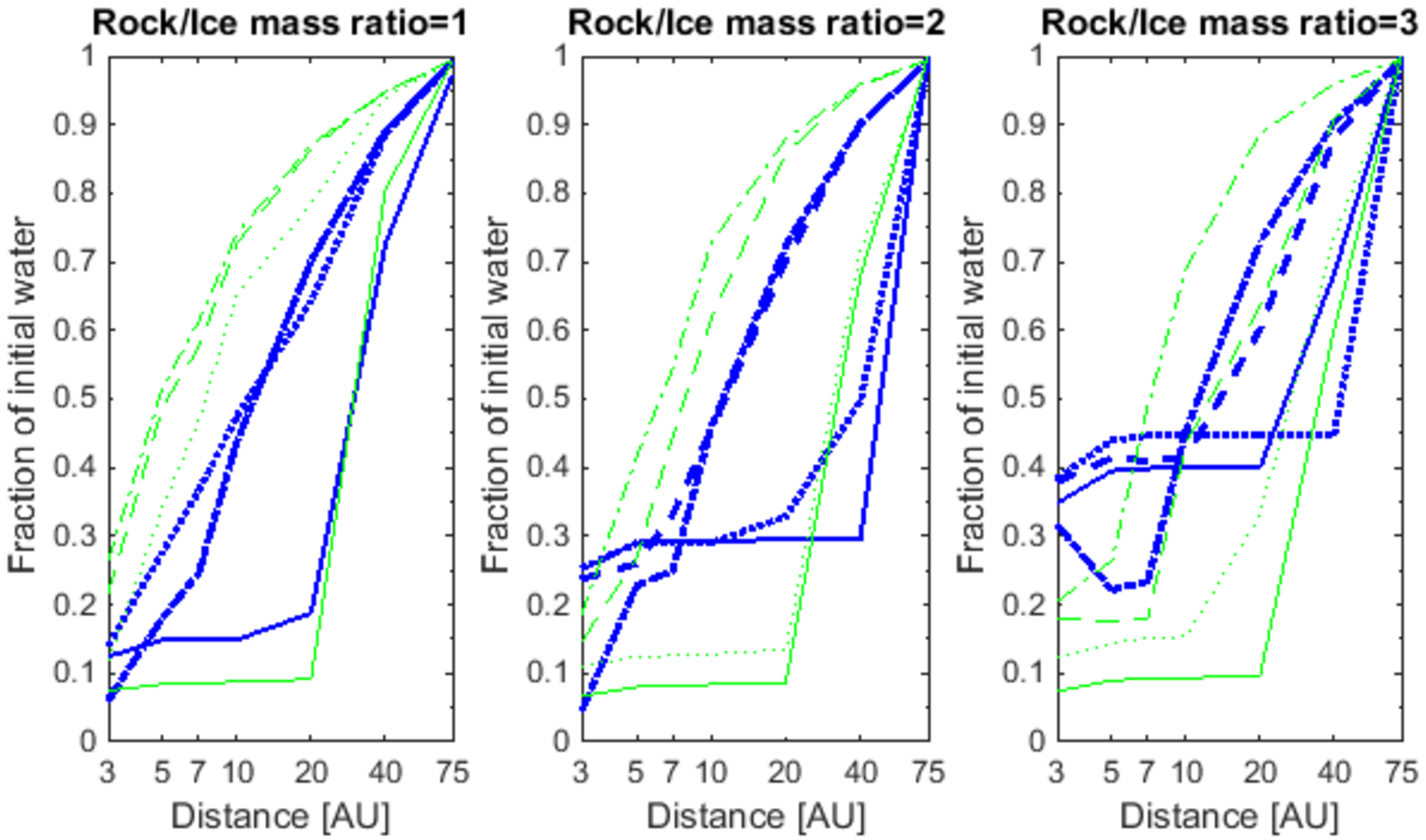}}
		\subfigure[2 M$_{\odot}$ progenitor - 0.59 M$_{\odot}$ WD - 1.34 Gyr evolution] {\label{fig:2sm}\includegraphics[scale=0.5]{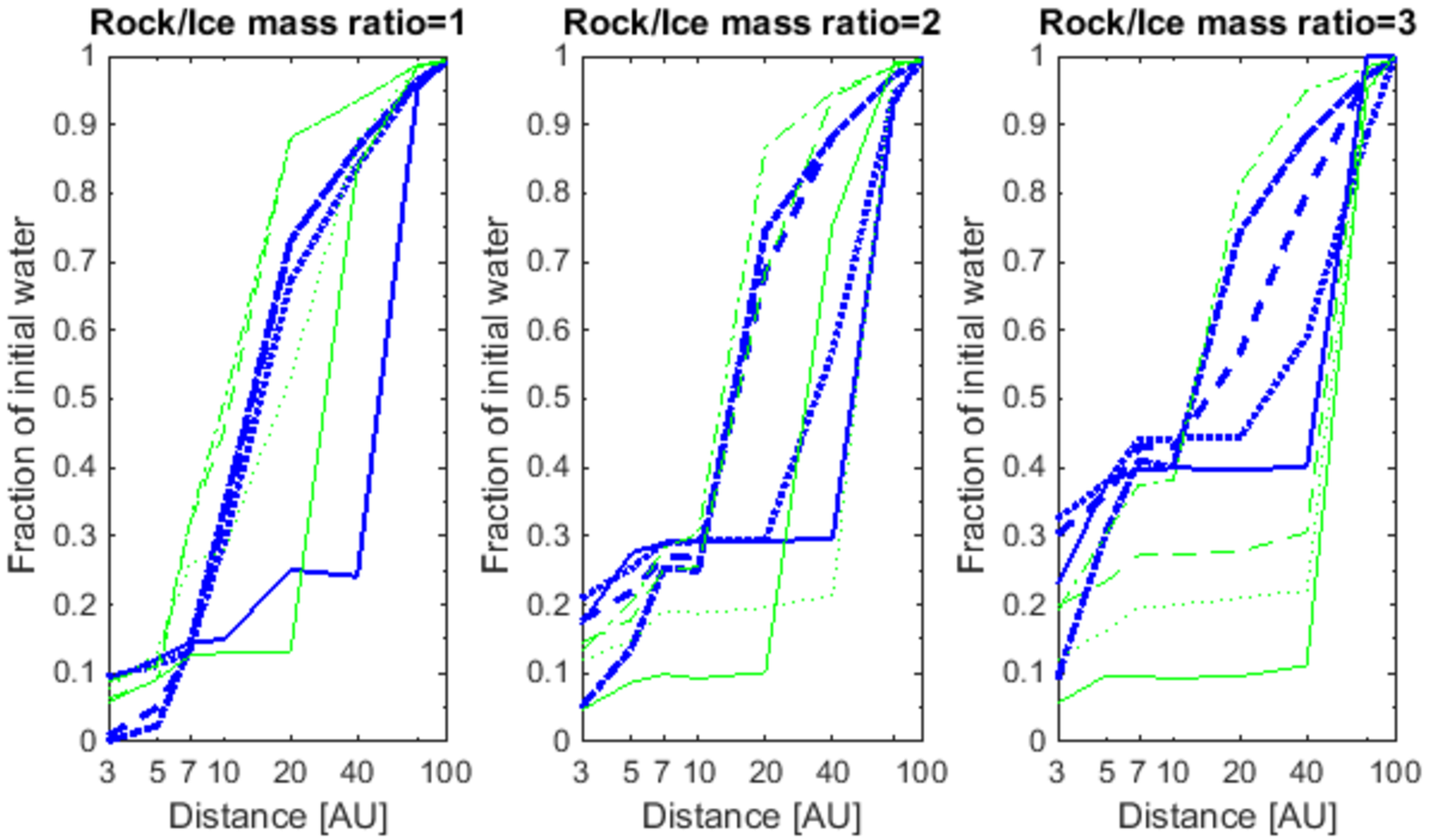}}
		\subfigure[3 M$_{\odot}$ progenitor - 0.65 M$_{\odot}$ WD - 472 Myr evolution] {\label{fig:3sm}\includegraphics[scale=0.5]{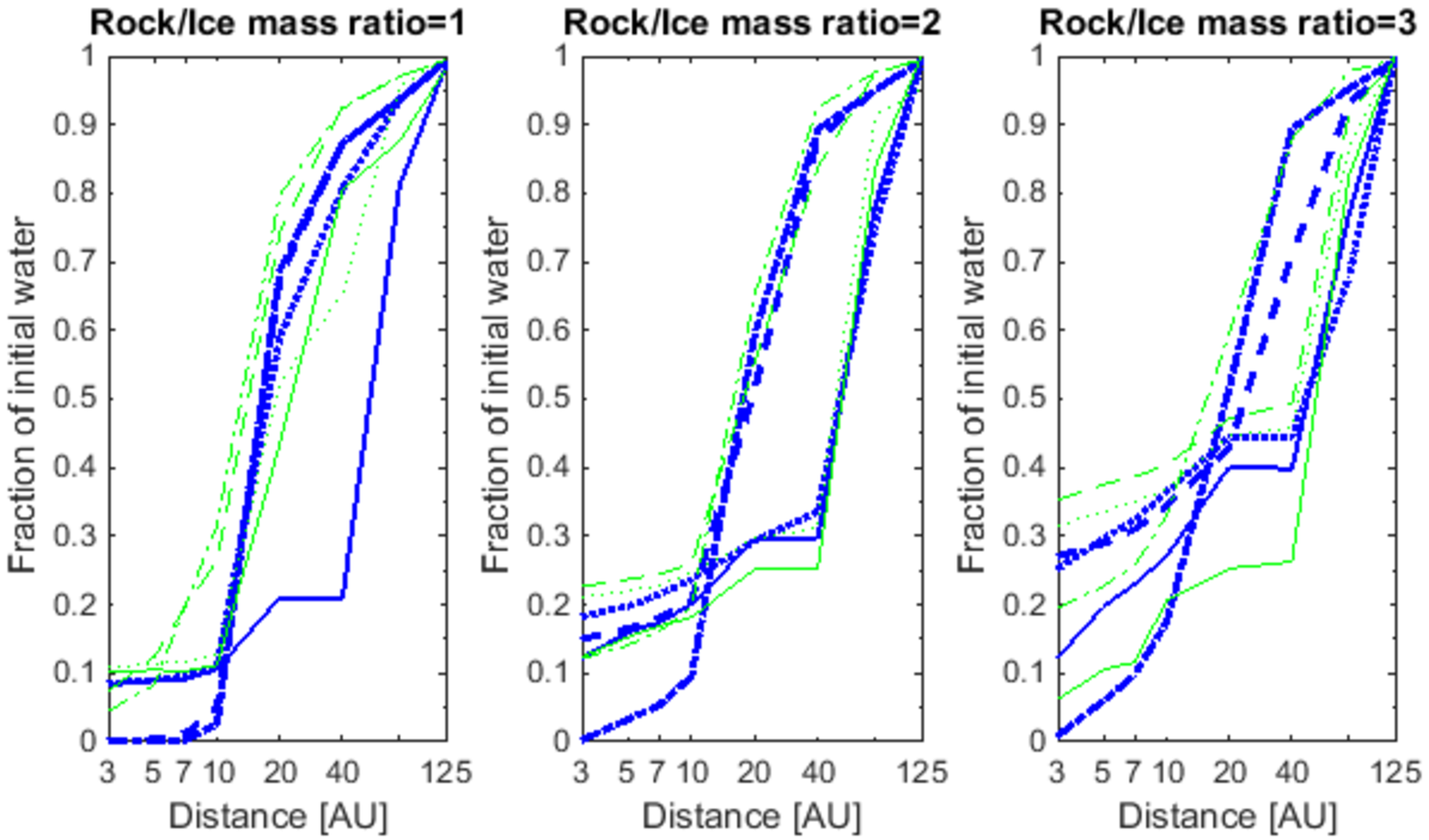}}
		\subfigure[3.6 M$_{\odot}$ progenitor - 0.76 M$_{\odot}$ WD - 278 Myr evolution] {\label{fig:36sm}\includegraphics[scale=0.5]{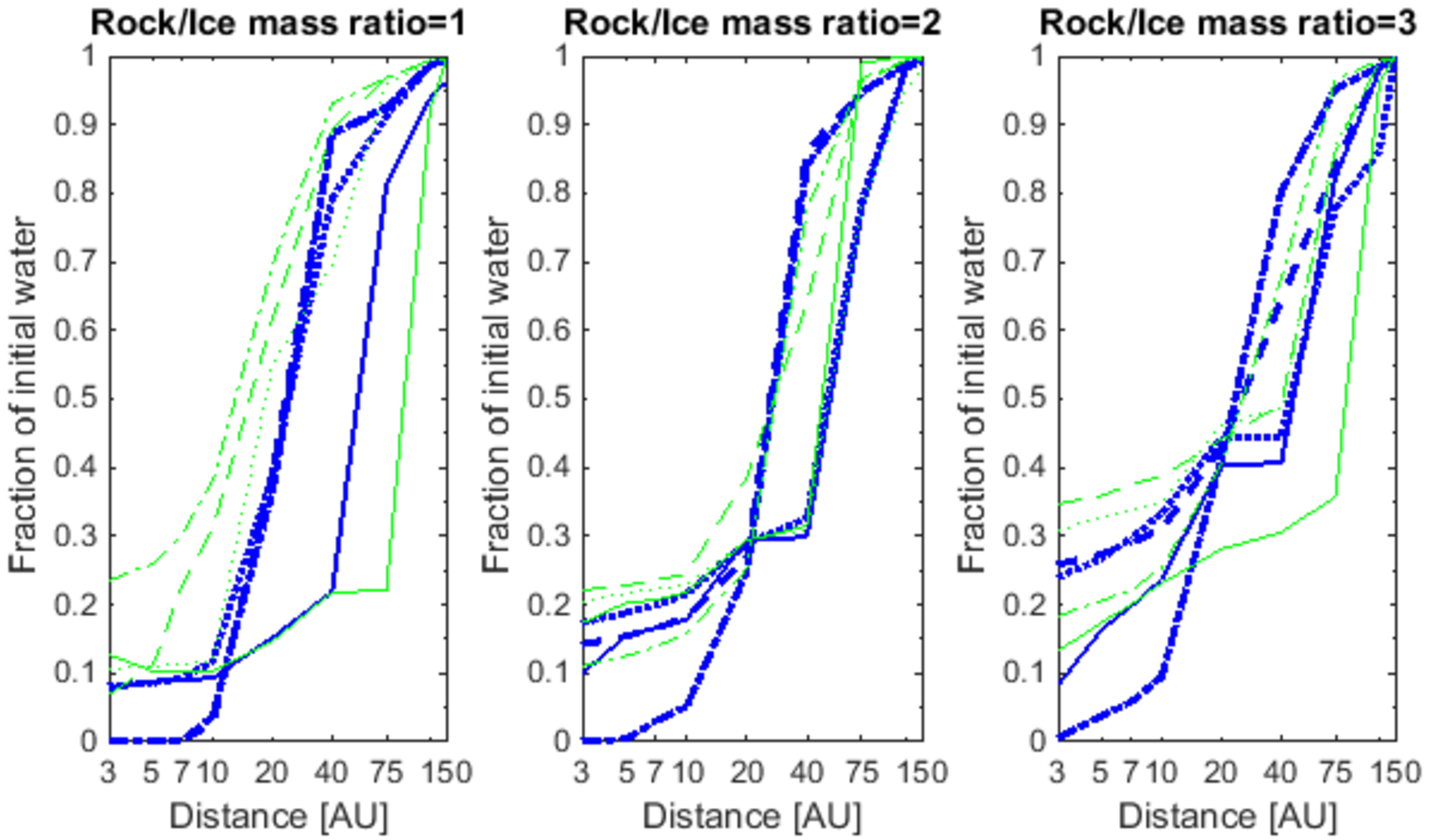}}
		\subfigure[5 M$_{\odot}$ progenitor - 0.89 M$_{\odot}$ WD - 117 Myr evolution] {\label{fig:5sm}\includegraphics[scale=0.5]{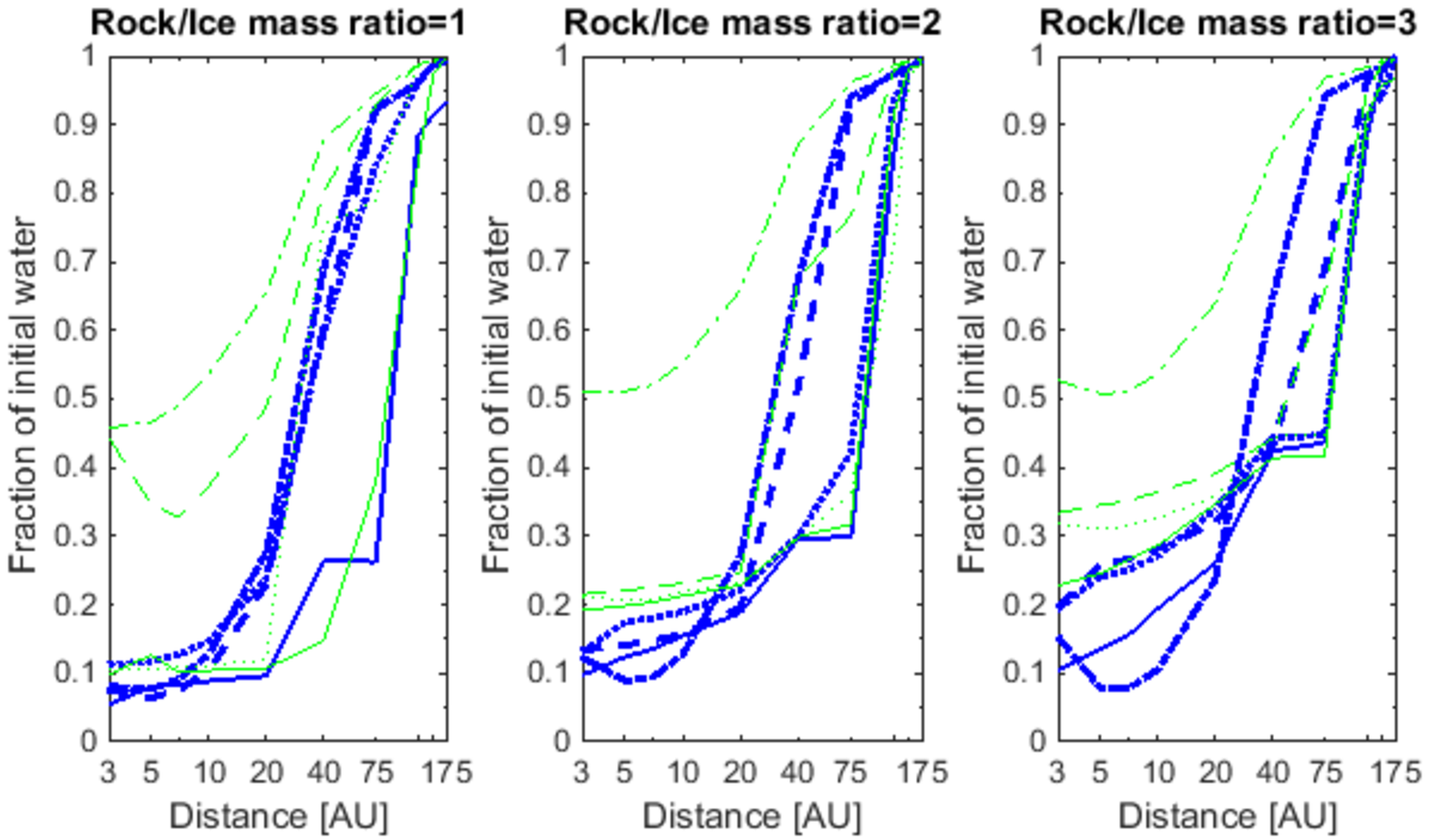}}
		\subfigure[6.4 M$_{\odot}$ progenitor - 1 M$_{\odot}$ WD - 65 Myr evolution] {\label{fig:64sm}\includegraphics[scale=0.5]{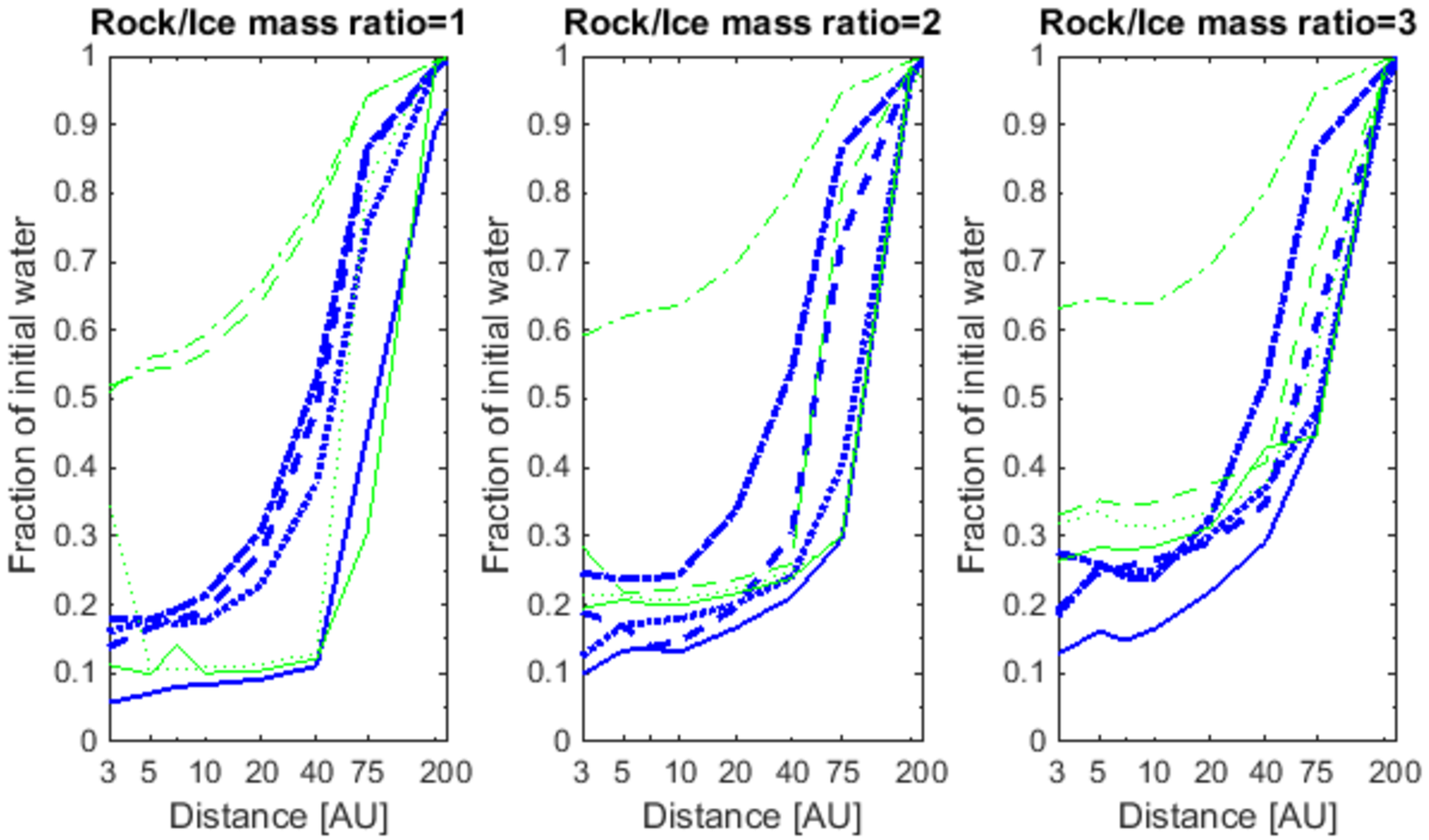}}
		\subfigure{\includegraphics[scale=0.5]{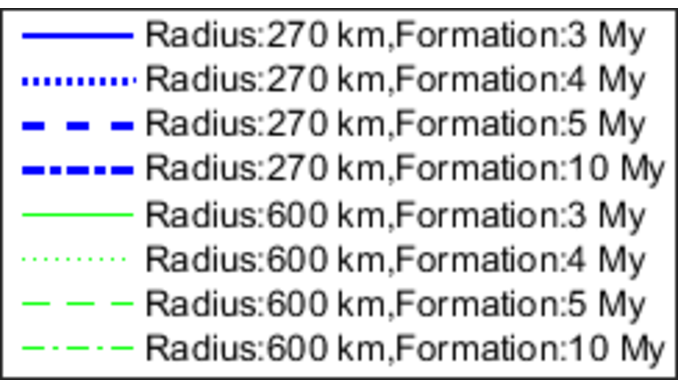}}
	\end{center}
	\caption{Total fraction of water (ice + water in hydrated silicates) remaining after the main sequence, RGB and AGB stellar evolution phases, for different progenitor masses with solar iron abundance [Fe/H]=0. The retention of water is shown as a function of the minor planet's initial orbital distance, composition, radius and formation time.}
	\label{fig:Masses}
\end{figure*}

In this section we discuss the bulk amount of water surviving in the planetary system, as a function of our free model parameters. Fig. \ref{fig:Masses} shows the final fraction of water, based on the end states of the production runs discussed in Section \ref{S:Model} (i.e., the ratio between the water when the star reaches the WD stage to the initial amount). We present the \emph{total} fraction of retained water, defined as water ice + water in hydrated silicates, which ultimately contributes hydrogen and oxygen when accreting onto polluted WD atmospheres. Each panel consists of three subplots, each representing a different choice for the initial composition. Within each subplot there are multiple lines, depicting the final water fraction as a function of the initial orbital distance. Each line is characterized by a specific color and width, as well as a style. The line width decreases with the size of the object, so the thin lines represent the more massive objects, and each line style corresponds to a different formation time.

Contrary to our previous work (MP17), where differences in progenitor star mass also entailed huge differences in water retention, here the differences appear to be much more subtle. The common general trend in the data is that minor planets at a greater distance from the star can better retain their water, as expected. There are two noticeable changes however, as the progenitor's mass increases. First, it can be seen that the outer bound of water retention increases with stellar mass. This is categorically true for any combination of parameters, and arises from the fact that stellar luminosity increases with star mass, and therefore the distance at which the minor planet's surface temperature can lead to sublimation extends outwards. It is also true for smaller minor planets, as reported by MP17.

The second, less trivial difference, is that at short orbital distances (less than $\sim$40 AU) around the least massive, 1 and 2 M$_{\odot}$ progenitors, the fraction of remaining water tends to be higher for minor planets with a 270 km radius compared to minor planets with a 600 km radius. This trend starts to change for 3 and 3.6 M$_{\odot}$ progenitors, and it reverses completely for 5 and 6.4 M$_{\odot}$ progenitors, where the more massive minor planets now retain a larger fraction of water. The reason for this phenomena is simply the relative fraction of rock dehydration of the outer layers. After all the ice is expelled from the minor planet, the outer layer temperatures begin to climb (primarily during the intense AGB phase), as we have seen in Section \ref{SS:Evolution}. Given a finite amount of time (for the star to reach the WD), the heat can only penetrate to a certain distance inward of the minor planet's surface and surpass the characteristic rock dehydration temperature, assuming of course that rock hydration temperatures were attained in the outer layers to begin with (which is not always the case if the formation time is too long, or rock/ice mass ratio too small). Smaller minor planets therefore tend to have a larger fraction of their outer rocks dehydrate, since the heat transport time is the same and the porosity profile is also similar, but their relative size is much smaller. On the other hand, with increasing progenitor mass, the stellar evolution time shortens dramatically. Thus the initial hydration of rock, becomes much more important than the subsequent dehydration of near-surface rock.

\section{Discussion and summary}\label{S:discussion}
In this study we investigate a wide range of progenitor star masses, relevant to G, F, A and B type stars. We also investigate dwarf-planet-sized icy objects, for the first time in any previous related water retention study. The results in Section \ref{SS:Retention} reaffirm the expectation that minor planets retain less water at closer distances to the star. However, the results differ from studies of smaller minor planets, since here the increase in the progenitor star's mass does not entail huge differences in water retention trends, primarily under 40 AU. Rather, the water retention in minor planets with the same assumed composition, follows a relatively similar trend regardless of the mass of the progenitor star. Beyond 40 AU, that is, at Kuiper belt distances, the progenitor's mass mainly changes the inclination of the slope, as the upper boundary of water retention extends outwards. 

Our conclusion is that most moderate-sized minor planets evolve similarly enough during their early evolution, that main-sequence and post-main-sequence sublimation effects are insufficient to set them apart significantly. Rather, their assumed initial parameters are more important. Particularly, minor planets in this study are large enough in order to attain rock hydration temperatures in nearly 100\% of the cases. This is why the fraction of water retention is almost always higher than zero. While at close orbital distances the ice completely sublimates, neither internal rock dehydration, nor external rock dehydration by the intense luminosity of massive progenitors seem to be able to change the outcome of higher than zero water retention.

It may thus be inferred that if the mass in the convection zone of certain polluted He-dominated WDs comes from the accretion event of a single moderate-sized minor planet, one might expect to always detect at least a small fraction of water (assuming the observation itself allows for such a detection). The detectable water fraction $f$ could be calculated by the multiplication of the initial assumed water fraction $f_w$ times the final fraction of water retention $f_r$. If we take the minimum initial orbital distance of 3 AU, and we average crudely over various progenitor masses, as well as object radii and formation times, we can qualitatively compute, as per assumed composition: (a) f=0.5$\times$$\sim$0.1=5\% when the rock/ice mass ratio=1; (b) f=0.33$\times$$\sim$0.15=5\% when the rock/ice mass ratio=2; and (c) f=0.25$\times$$\sim$0.2=5\% when the rock/ice mass ratio=3. In other words, irrespective of progenitor mass and other parameters, the water fraction in single, moderate-sized exo-planetary minor planets should be of the order of 5\% even if their initial orbital distance is assumed to be 3 AU. The inverse reasoning may also be applied. If it indeed turns out that observationally, most polluted WD atmospheres are much dryer than about 5\%, it may perhaps be inferred that the accreted mass typically arises from multiple small accretion events and not singular large accretion events.

An additional interesting result is related to the question of habitability. A minor planet, or perhaps a planet accompanied by its moderate-sized moon/s, could be orbiting at $\sim$0.01 AU around a non-magnetic, relatively cool WD, and thus be potentially habitable for approximately 3-8 Gyr, providing ample time for life to develop \citep{Agol-2011,FossatiEtAl-2012}. Our results would suggest that while small minor planets may not always retain sufficient water to support life, especially if they are perturbed from close heliocentric distances, moderate-sized minor planets may have a far greater chance to do so.

This paper finalizes a series of papers on WD pollution and water retention (following MP16 and MP17). The results presented in MP17 for small minor planets, are complemented here with results for massive minor planets. Together they span nearly the entire potential minor planet mass range, as inferred from WD polluted atmospheres. We thus provide, as a service to the community, a code which may be used in order to evaluate water retention, covering the full mass range, of both minor planets and their respective host stars. The code runs on MATLAB, and is freely available at \url{https://github.com/UriMalamud/WaterRetention}.

\section{Acknowledgment}\label{S:Acknowledgment}
UM and HBP acknowledge support from the Marie Curie FP7 career integration grant "GRAND", the Research Cooperation Lower Saxony-Israel Niedersachsisches Vorab fund, the Minerva center for life under extreme planetary conditions and the ISF I-CORE grant 1829/12.

\newpage


\bibliographystyle{apj} 

\end{document}